\documentstyle[prl,aps,preprint]{revtex}
\begin{document}
\bigskip\ 

\begin{center}
{\bf S-DUALITY FOR LINEARIZED GRAVITY}

\bigskip\ 

\smallskip\ 

J. A. Nieto\footnote[1]{%
nieto@uas.uasnet.mx}

\smallskip\ 

{\it Facultad de Ciencias F\'{\i}sico-Matem\'{a}ticas de la Universidad
Aut\'{o}noma}

{\it de Sinaloa, 80010 Culiac\'{a}n Sinaloa, M\'{e}xico}

\bigskip\ 

\bigskip\ 

{\bf Abstract}
\end{center}

We develope the analogue of S-duality for linearized gravity in
(3+1)-dimensions. Our basic idea is to consider the self-dual
(anti-self-dual) curvature tensor for linearized gravity in the context of
the Macdowell-Mansouri formalism. We find that the strong-weak coupling
duality for linearized gravity is an exact symmetry and implies small-large
duality for the cosmological constant.

\bigskip\ 

Pacs numbers: 04.60.-m, 04.65.+e, 11.15.-q, 11.30.Ly

August, 1999

\newpage\ 

\smallskip\ 

\noindent {\bf 1.- INTRODUCTION}

\smallskip\ 

It is well known that linearized Einstein gravitational theory in four
dimensions is similar to the source free Maxwell theory. Since S-duality has
been successfully applied to Abelian gauge theories and non-Abelian gauge
theories [1-10] as well as gravity [11-15], it seems natural to ask whether
similar methods may be applied to linearized gravity. However, while in the
Abelian case there is an exact strong-weak duality, Yang-Mills and gravity
do not possess such a symmetry.

In this work, we show that in spite of gravity, in general, does not possess
an exact strong-weak symmetry linearized gravity does. In fact, we derive a
S-dual action for linearized gravity which is connected to the original
linearized gravitational action by strong-weak duality transformation. Our
strategy essentially consists in two steps; first in writing linearized
gravity as Abelian gauge theory and second taking recourse of the
Macdowell-Mansouri formalism.

Just as $U(1)$ duality has been very important to explore S-duality in
larger theories such as $N=2$ super Yang-Mills theory with gauge group $%
SU(2) $ [16-17]$,$ one should expect that linearized gravitational duality
may also be an important tool for investigating S-duality analogue of other
theories such as $N=1$ supergravity in four dimensions.

The plan of this work is as follows: In section II, we show that it is
possible to see linearized gravity as an Abelian gauge theory. In section
III, we briefly review the Macdowell-Mansouri theory and propose an action
for linearized gravity in the context of such a theory. In section IV, we
generalize our proposed action to include self-dual and anti-self-dual
linearized gravity and we compute the partition function of such a
generalized action. Finally, in section V, we make some final comments.

\smallskip\ 

\noindent {\bf 2.- LINEARIZED GRAVITY AS A GAUGE THEORY}

\smallskip\ 

The starting point in linearized gravity is to write the metric of the
space-time $g_{\mu \nu }=g_{\mu \nu }(x^\alpha )$ as 
\begin{equation}
g_{\mu \nu }=\eta _{\mu \nu }+h_{\mu \nu },  \label{1}
\end{equation}
where $h_{\mu \nu }$ is a small deviation of the metric $g_{\mu \nu }$ from
the Minkowski metric 
\[
(\eta _{\mu \nu })=diag(-1,1,1,1). 
\]

The first-order curvature tensor is 
\begin{equation}
F_{\mu \nu \alpha \beta }=\frac 12(\partial _\mu \partial _\alpha h_{\nu
\beta }-\partial _\mu \partial _\beta h_{\nu \alpha }-\partial _\nu \partial
_\alpha h_{\mu \beta }+\partial _\nu \partial _\beta h_{\mu \alpha }),
\label{2}
\end{equation}
which is invariant under the gauge transformation 
\begin{equation}
\delta h_{\mu \nu }=\partial _\mu \xi _\nu +\partial _\nu \xi _\mu ,
\label{3}
\end{equation}
Here, $\xi _\mu $ is an arbitrary vector field and $\partial _\mu \equiv 
\frac \partial {\partial x^\mu }$.

It is not difficult to see that $F_{\mu \nu \alpha \beta }$ satisfies the
following relations: 
\begin{equation}
\begin{array}{c}
F_{\mu \nu \alpha \beta }=-F_{\mu \nu \beta \alpha }=-F_{\nu \mu \alpha
\beta }=F_{\alpha \beta \mu \nu }, \\ 
\\ 
F_{\mu \nu \alpha \beta }+F_{\mu \beta \nu \alpha }+F_{\mu \alpha \beta \nu
}=0, \\ 
\\ 
\partial _\lambda F_{\mu \nu \alpha \beta }+\partial _\mu F_{\nu \lambda
\alpha \beta }+\partial _\nu F_{\lambda \mu \alpha \beta }=0.
\end{array}
\label{4}
\end{equation}
It is interesting to note that the dual $^{*}F_{\mu \nu \alpha \beta }\equiv 
\frac 12\epsilon _{\mu \nu \tau \sigma }F_{\alpha \beta }^{\tau \sigma },$
where $\epsilon _{\mu \nu \alpha \beta }$ is a completely antisymmetric
Levi-Civita tensor with $\epsilon _{0123}=-1$ and the indices are raised and
lowered by means of $\eta ^{\alpha \beta }$ and $\eta _{\alpha \beta },$
does not satisfy the relations (4) unless the vacuum Einstein equations $%
F_{\mu \nu }=F_{\mu \alpha \nu }^\alpha =0$ are satisfied [18].

Let us introduce the `gauge' field 
\begin{equation}
A_{\mu \alpha \beta }=\frac 12(\partial _\alpha h_{\mu \beta }-\partial
_\beta h_{\mu \alpha }).  \label{5}
\end{equation}
Note that $A_{\mu \alpha \beta }=-A_{\mu \beta \alpha }.$ Using (3), we find
that $A_{\mu \alpha \beta }$ transforms as 
\begin{equation}
\delta A_{\mu \alpha \beta }=\partial _\mu \lambda _{\alpha \beta },
\label{6}
\end{equation}
where $\lambda _{\alpha \beta }=\partial _\alpha \xi _\beta -\partial _\beta
\xi _\alpha $.

The expression (5) allows to write the curvature tensor $F_{\mu \nu
}^{\alpha \beta }$ as 
\begin{equation}
F_{\mu \nu }^{\alpha \beta }=\partial _\mu A_\nu ^{\alpha \beta }-\partial
_\nu A_\mu ^{\alpha \beta }.  \label{7}
\end{equation}
Thus, we have shown that the tensor $F_{\mu \nu }^{\alpha \beta }$ may be
written in the typical form of an abelian Maxwell field strength $F_{\mu \nu
}^a=\partial _\mu A_\nu ^a-\partial _\nu A_\mu ^a,$ where the index $a$ runs
over some abelian group such as $U(1)\times U(1)\times ...\times U(1)$. Note
that $F_{\mu \nu }^{\alpha \beta }$ is invariant under the transformation $%
\delta A_\mu ^{\alpha \beta }=\partial _\mu \lambda ^{\alpha \beta }$ which
has exactly the same form as the transformation of Abelian gauge fields $%
\delta A_\mu ^a=\partial _\mu \lambda ^a,$ where $\lambda ^a$ is an
arbitrary function of the coordinates $x^\mu $.

\smallskip\ 

\noindent {\bf 3.- LINEARIZED GRAVITY a l\'{a} MACDOWELL-MANSOURI}

\smallskip\ 

The central idea in the Macdowell-Mansouri theory [19-20] (see also Refs.
[21-31]) is to represent the gravitational field as a connection one-form
associated to some group that contains the Lorentz group as a subgroup. The
typical example is provided by the SO(3,2) anti-de Sitter gauge theory of
gravity. In this case the SO(3,2) gravitational gauge field $\omega _\mu
^{AB}=-$ $\omega _\mu ^{BA}$ is broken into the SO(3,1) connection $\omega
_\mu ^{ab}$ and the $\omega _\mu ^{4a}=e_\mu ^a$ vierbein field. Thus, the
anti-de Sitter curvature 
\begin{equation}
{\cal R}_{\mu \nu }^{AB}=\partial _\mu \omega _\nu ^{AB}-\partial _\nu
\omega _\mu ^{AB}+\omega _\mu ^{AC}\omega _{\nu C}^{\quad B}-\omega _\nu
^{AC}\omega _{\mu C}^{\quad B}  \label{8}
\end{equation}
leads to 
\begin{equation}
{\cal R}_{\mu \nu }^{ab}=R_{\mu \nu }^{ab}+\Sigma _{\mu \nu }^{ab}  \label{9}
\end{equation}
and 
\begin{equation}
{\cal R}_{\mu \nu }^{4a}=\partial _\mu e_\nu ^a-\partial _\nu e_\mu
^a+\omega _\mu ^{ac}e_{\nu c}-\omega _\nu ^{ac}e_{\mu c},  \label{10}
\end{equation}
where 
\begin{equation}
R_{\mu \nu }^{ab}=\partial _\mu \omega _\nu ^{ab}-\partial _\nu \omega _\mu
^{ab}+\omega _\mu ^{ac}\omega _{\nu c}^{\quad b}-\omega _\nu ^{ac}\omega
_{\mu c}^{\quad b}  \label{11}
\end{equation}
is the SO(3,1) curvature and 
\begin{equation}
\Sigma _{\mu \nu }^{ab}=e_\mu ^ae_\nu ^b-e_\nu ^ae_\mu ^b.  \label{12}
\end{equation}
It turns out that $T_{\mu \nu }^a\equiv {\cal R}_{\mu \nu }^{4a}$ can be
identified with the torsion.

The Macdowell-Mansouri's action is 
\begin{equation}
S=\frac 14\int d^4x{}\varepsilon ^{\mu \nu \alpha \beta }{}{\cal R}_{\mu \nu
}^{ab}{}{\cal R}_{\alpha \beta }^{cd}{}\epsilon _{abcd},  \label{13}
\end{equation}
where $\varepsilon ^{\mu \nu \alpha \beta }$ is the completely antisymmetric
tensor associated to the space-time, with $\varepsilon ^{0123}=1$ and $%
\varepsilon _{0123}=1$, while $\epsilon _{abcd}$ is also the completely
antisymmetric tensor but now associated to the internal group S(3,1), with $%
\epsilon _{0123}=-1.$ We assume that the internal metric is given by $(\eta
_{ab})=(-1,1,1,1).$ Therefore, we have $\epsilon ^{0123}=1.$ It is well
known that, using (9), (11) and (12), the action (13) leads to three terms;
the Hilbert Einstein action, the cosmological constant term and the Euler
topological invariant (or Gauss-Bonnet term). It is worth mentioning that
the action (13) may also be obtained using Lovelock theory (see [32] and
references there in).

Let us now apply the Macdowell-Mansouri formalism to linearized gravity as
developed in section 2.

First, consider the extended curvature 
\begin{equation}
{\cal F}_{\mu \nu }^{\alpha \beta }=F_{\mu \nu }^{\alpha \beta }+\Omega
_{\mu \nu }^{\alpha \beta },  \label{14}
\end{equation}
where

\begin{equation}
\Omega _{\mu \nu }^{\alpha \beta }=\delta _\mu ^\alpha h_\nu ^\beta -\delta
_\mu ^\beta h_\nu ^\alpha -\delta _\nu ^\alpha h_\mu ^\beta +\delta _\nu
^\beta h_\mu ^\alpha .  \label{15}
\end{equation}
The central idea is to consider the extended curvature (14) as the analogue
of the curvature (9) of the Macdowell-Mansouri formalism. Thus, we may
identify $F_{\mu \nu }^{\alpha \beta }$ and $\Omega _{\mu \nu }^{\alpha
\beta }$ with the linearized versions of $R_{\mu \nu }^{ab}$ and $\Sigma
_{\mu \nu }^{ab},$ respectively. In fact, if we write $e_\mu ^a=b_\mu
^a+l_\mu ^a$, where $b_\mu ^a$ corresponds to the flat metric $\eta _{\mu
\nu }$ and $l_\mu ^a$ is a small deviation of $e_\mu ^a$ , we find $h_{\mu
\nu }=b_\mu ^al_\nu ^b\eta _{ab}+b_\nu ^al_\mu ^b\eta _{ab}$ and therefore
using (5) we get $A_{\mu \alpha \beta }=\omega _{\mu \alpha \beta
}(l_\lambda ^a)+\partial _\mu f_{\alpha \beta }.$ Here, $\omega _{\mu \alpha
\beta }(l_\lambda ^a)=b_{a\alpha }b_{b\beta }\omega _\mu ^{ab}(l_\lambda ^a)$
and $f_{\alpha \beta }=b_{a\alpha }l_\beta ^a-b_{a\beta }l_\alpha ^a.$ (It
is important to note that $\omega _\mu ^{ab}(l_\lambda ^a)$ can be obtained
by setting the torsion (10), for weak gravity, equal to zero.) Thus, up to a
gauge transformation $A_{\mu \alpha \beta }$ is equal to the linearized
connection $\omega _\mu ^{ab}(l_\lambda ^a).$ The essential difference
between $\omega _\mu ^{ab}(l_\lambda ^a)$ and the full connection $\omega
_\mu ^{ab}$ is that the latter is associated to the Lorentz group while the
former to an abelian group. This can easily be seen from the transformation $%
\delta A_\mu ^{\alpha \beta }=\partial _\mu \lambda ^{\alpha \beta }$ of the
gauge field $A_\mu ^{\alpha \beta },$ since in general a gauge field $A$
transforms as $\delta A=\partial \lambda +[A,\lambda ].$

Let us now propose the action 
\begin{equation}
S=\frac 14\int d^4x{}\epsilon ^{\mu \nu \alpha \beta }{}{\cal F}_{\mu \nu
}^{\tau \lambda }{}{\cal F}_{\alpha \beta }^{\sigma \rho }{}\epsilon _{\tau
\lambda \sigma \rho }.  \label{16}
\end{equation}
We shall show that this action is reduced to Einstein linearized
gravitational action (Fierz-Pauli action ) with cosmological constant and a
total derivative (`topological' term). Substituting (14) into (16) yields 
\begin{equation}
\begin{array}{c}
S=\frac 14\int d^4x{}\epsilon ^{\mu \nu \alpha \beta }{}F_{\mu \nu }^{\tau
\lambda }{}F_{\alpha \beta }^{\sigma \rho }{}\epsilon _{\tau \lambda \sigma
\rho }+\frac 12\int d^4x{}\epsilon ^{\mu \nu \alpha \beta }{}\Omega _{\mu
\nu }^{\tau \lambda }{}F_{\alpha \beta }^{\sigma \rho }{}\epsilon _{\tau
\lambda \sigma \rho } \\ 
\\ 
+\frac 14\int d^4x{}\epsilon ^{\mu \nu \alpha \beta }{}\Omega _{\mu \nu
}^{\tau \lambda }{}\Omega _{\alpha \beta }^{\sigma \rho }{}\epsilon _{\tau
\lambda \sigma \rho }.
\end{array}
\label{17}
\end{equation}
Using (15) we find that (17) is reduced to 
\begin{equation}
\begin{array}{c}
S=\frac 14\int d^4x{}\epsilon ^{\mu \nu \alpha \beta }{}F_{\mu \nu }^{\tau
\lambda }{}F_{\alpha \beta }^{\sigma \rho }{}\epsilon _{\tau \lambda \sigma
\rho }+2\int d^4x{}\epsilon ^{\mu \nu \alpha \beta }{}\delta _\mu ^\tau
{}h_\nu ^\lambda F_{\alpha \beta }^{\sigma \rho }{}\epsilon _{\tau \lambda
\sigma \rho } \\ 
\\ 
+4\int d^4x{}\epsilon ^{\mu \nu \alpha \beta }{}\delta _\mu ^\tau h_\nu
^\lambda {}\delta _\alpha ^\sigma h_\beta ^\rho {}\epsilon _{\tau \lambda
\sigma \rho }.
\end{array}
\label{18}
\end{equation}
Since $\epsilon ^{\mu \nu \alpha \beta }{}\delta _\mu ^\tau {}\epsilon
_{\tau \lambda \sigma \rho }=-\delta _{\lambda \sigma \rho }^{\nu \alpha
\beta }$ and $\epsilon ^{\mu \nu \alpha \beta }{}\delta _\mu ^\tau {}\delta
_\alpha ^\sigma {}\epsilon _{\tau \lambda \sigma \rho }=-2{}\delta _{\lambda
\rho }^{\nu \beta }$, where in general $\delta _{\tau \lambda \sigma \rho
}^{\mu \nu \alpha \beta }$ is a generalized delta, we discover that (18) can
be written as 
\begin{equation}
\begin{array}{c}
S=\frac 14\int d^4x{}\epsilon ^{\mu \nu \alpha \beta }{}F_{\mu \nu }^{\tau
\lambda }{}F_{\alpha \beta }^{\sigma \rho }{}\epsilon _{\tau \lambda \sigma
\rho }+8\int d^4x{}h^{\mu \nu }(F_{\mu \nu }-\frac 12\eta _{\mu \nu }F) \\ 
\\ 
-8\int d^4x{}(h^2-h^{\mu \nu }h_{\mu \nu }).
\end{array}
\label{19}
\end{equation}
Here, we used the following definitions: $F_{\mu \nu }\equiv \eta ^{\alpha
\beta }F_{\mu \alpha \nu \beta },F\equiv \eta ^{\mu \nu }\eta ^{\alpha \beta
}F_{\mu \alpha \nu \beta }$ and $h=\eta ^{\mu \nu }h_{\mu \nu }.$ We
recognize the second term and the third term in (19) as the Einstein action
for linearized gravity with cosmological constant, while the first term is a
total derivative (an Euler topological invariant or Gauss-Bonnet term). Note
that the usual cosmological factor $\Lambda $ in the third term can be
derived simply by changing $\Omega \rightarrow a^2\Omega ,$ where $a$ is a
constant, and rescaling the total action $S\rightarrow \frac 14\Lambda
^{-1}S,$ with $\Lambda =a^2.$

\smallskip\ 

\noindent {\bf 4.- S-DUALITY FOR LINEARIZED GRAVITY}

\smallskip\ 

In order to develope a S-dual linearized gravitational action we generalize
the action (16) as follows; 
\begin{equation}
{\cal S}=\frac 14(\tau ^{+})\int d^4x{}\epsilon ^{\mu \nu \alpha \beta
}{}^{+}{\cal F}_{\mu \nu }^{\tau \lambda }{}^{+}{\cal F}_{\alpha \beta
}^{\sigma \rho }{}\epsilon _{\tau \lambda \sigma \rho }-\frac 14(\tau
^{-})\int d^4x{}\epsilon ^{\mu \nu \alpha \beta }{}^{-}{\cal F}_{\mu \nu
}^{\tau \lambda }{}^{-}{\cal F}_{\alpha \beta }^{\sigma \rho }{}\epsilon
_{\tau \lambda \sigma \rho },  \label{20}
\end{equation}
where $\tau ^{+}$ and $\tau ^{-}$ are two different constant parameters and $%
^{\pm }{\cal F}_{\mu \nu }^{\alpha \beta }$ is given by 
\begin{equation}
^{\pm }{\cal F}_{\mu \nu }^{\alpha \beta }=(\frac 12)^{\pm }M_{\tau \lambda
}^{\alpha \beta }{\cal F}_{\mu \nu }^{\tau \lambda },  \label{21}
\end{equation}
where 
\begin{equation}
^{\pm }M_{\tau \lambda }^{\alpha \beta }=\frac 12(\delta _{\tau \lambda
}^{\alpha \beta }\mp i\epsilon _{\,\quad \tau \lambda }^{\alpha \beta }).
\label{22}
\end{equation}
It turns out that $^{+}{\cal F}_{\mu \nu }^{\alpha \beta }$ is self-dual,
while $^{-}{\cal F}_{\mu \nu }^{\alpha \beta }$ is anti self-dual curvature.
Therefore, the action (20) describes self-dual and anti-self-dual linearized
gravity.

We may now follow a similar procedure as in the references [5-8] for Abelian
Yang-Mills and [11, 14, 15] for gravity. Here, however, in order to show the
exactness of S-duality symmetry in linearized gravity, it turns out to be
more convenient to follow the procedure of reference [1] due to Witten. For
this purpose let us first introduce a two form $G$ and let us set 
\begin{equation}
{\cal H}_{\mu \nu }^{\alpha \beta }\equiv {\cal F}_{\mu \nu }^{\alpha \beta
}-G_{\mu \nu }^{\alpha \beta }.  \label{23}
\end{equation}
We assume that $G_{\mu \nu }^{\alpha \beta }$ satisfies the same symmetric
properties as $F_{\mu \nu }^{\alpha \beta },$ given in (4), namely 
\begin{equation}
\begin{array}{c}
G_{\mu \nu \alpha \beta }=-G_{\mu \nu \beta \alpha }=-G_{\nu \mu \alpha
\beta }=G_{\alpha \beta \mu \nu }, \\ 
\\ 
G_{\mu \nu \alpha \beta }+G_{\mu \beta \nu \alpha }+G_{\mu \alpha \beta \nu
}=0.
\end{array}
\label{24}
\end{equation}

Now, consider the extended action 
\begin{equation}
\begin{array}{ccc}
{\cal S}_E= & \frac 14(\tau ^{+})\int dx^4\varepsilon ^{\mu \nu \alpha \beta
}{}^{+}{\cal H}_{\mu \nu }^{\tau \lambda }{}^{+}{\cal H}_{\alpha \beta
}^{\sigma \rho }{}\epsilon _{\tau \lambda \sigma \rho }-\frac 14(\tau
^{-})\int dx^4\varepsilon ^{\mu \nu \alpha \beta }{}^{-}{\cal H0}_{\mu \nu
}^{\tau \lambda }{}^{-}{\cal H}_{\alpha \beta }^{\sigma \rho }{}\epsilon
_{\tau \lambda \sigma \rho } &  \\ 
&  &  \\ 
& +\frac 12\int d^4x\varepsilon ^{\mu \nu \tau \lambda }{}^{+}W_{\mu \nu
}^{\alpha \beta }{}^{+}G_{\tau \lambda }^{\sigma \rho }{}\epsilon _{\alpha
\beta \sigma \rho }-\frac 12\int d^4x\varepsilon ^{\mu \nu \tau \lambda
}{}^{-}W_{\mu \nu }^{\alpha \beta }{}^{-}G_{\tau \lambda }^{\sigma \rho
}{}\epsilon _{\alpha \beta \sigma \rho }, & 
\end{array}
\label{25}
\end{equation}
where $W_{\mu \nu \alpha \beta }=\partial _\mu V_{\nu \alpha \beta
}-\partial _\nu V_{\mu \alpha \beta }$ is the dual field strength satisfying
the Dirac quatization law 
\begin{equation}
\int W\in 2\pi {\bf Z.}  \label{26}
\end{equation}
It is not difficult to see that, beyond the gauge invariance $A\rightarrow
A-d\lambda ,$ $G\rightarrow G$, the partition function

\begin{equation}
Z=\int d^{+}G{}d^{-}G{}dA{}dh{}dV{}e^{-{\cal S}_E}  \label{27}
\end{equation}
is invariant under 
\begin{equation}
A\rightarrow A+B\;and\;G\rightarrow G+dB,  \label{28}
\end{equation}
where $B_\mu ^{\alpha \beta }$ is an arbitrary one form.

Starting from (25) one can proceed in two different ways. For the first
possibility, we note that the path integral that involves $V$ is 
\begin{equation}
\int DV\exp (\frac 12\int d^4x\varepsilon ^{\mu \nu \tau \lambda
}{}^{+}W_{\mu \nu }^{\alpha \beta }{}^{+}G_{\tau \lambda }^{\sigma \rho
}{}\epsilon _{\alpha \beta \sigma \rho }-\frac 12\int d^4x\varepsilon ^{\mu
\nu \tau \lambda }{}^{-}W_{\mu \nu }^{\alpha \beta }{}^{-}G_{\tau \lambda
}^{\sigma \rho }{}\epsilon _{\alpha \beta \sigma \rho }).  \label{29}
\end{equation}
Integrating over the dual connection $V$, we get a delta function setting $%
dG=0.$ Thus, using the gauge invariance (28), we may gauge $G$ to zero,
reducing (25) to the original action (20). Therefore, the actions (25) and
(20) are, in fact, classically equivalents.

For the second possibility, we note that the gauge invariance (28) enables
one to fix a gauge with $A=0.$ (It is important to note that, at this stage,
we are considering $A_{\mu \alpha \beta }$ and $h_{\mu \nu }$ as independent
fields in the sense of Palatini.) The action (25) is then reduced to 
\begin{equation}
\begin{array}{ccc}
{\cal S}_E= & \frac 14(\tau ^{+})\int dx^4\varepsilon ^{\mu \nu \alpha \beta
}{}^{+}{\cal G}_{\mu \nu }^{\tau \lambda }{}^{+}{\cal G}_{\alpha \beta
}^{\sigma \rho }{}\epsilon _{\tau \lambda \sigma \rho }-\frac 14(\tau
^{-})\int dx^4\varepsilon ^{\mu \nu \alpha \beta }{}^{-}{\cal G}_{\mu \nu
}^{\tau \lambda }{}^{-}{\cal G}_{\alpha \beta }^{\sigma \rho }{}\epsilon
_{\tau \lambda \sigma \rho } &  \\ 
&  &  \\ 
& +\frac 12\int d^4x\varepsilon ^{\mu \nu \tau \lambda }{}^{+}W_{\mu \nu
}^{\alpha \beta }{}^{+}G_{\tau \lambda }^{\sigma \rho }{}\epsilon _{\alpha
\beta \sigma \rho }-\frac 12\int d^4x\varepsilon ^{\mu \nu \tau \lambda
}{}^{-}W_{\mu \nu }^{\alpha \beta }{}^{-}G_{\tau \lambda }^{\sigma \rho
}{}\epsilon _{\alpha \beta \sigma \rho }, & 
\end{array}
\label{30}
\end{equation}
where 
\begin{equation}
{\cal G}_{\mu \nu }^{\tau \lambda }\equiv \Omega _{\mu \nu }^{\tau \lambda
}-G_{\mu \nu }^{\tau \lambda }.  \label{31}
\end{equation}
In virtue of (31) the extended action (30) becomes 
\begin{equation}
\begin{array}{ccc}
{\cal S}_E= & \frac 14(\tau ^{+})\int dx^4\varepsilon ^{\mu \nu \alpha \beta
}{}^{+}G_{\mu \nu }^{\tau \lambda }{}^{+}G_{\alpha \beta }^{\sigma \rho
}{}\epsilon _{\tau \lambda \sigma \rho }-\frac 14(\tau ^{-})\int
dx^4\varepsilon ^{\mu \nu \alpha \beta }{}^{-}G_{\mu \nu }^{\tau \lambda
}{}^{-}G_{\alpha \beta }^{\sigma \rho }{}\epsilon _{\tau \lambda \sigma \rho
} &  \\ 
&  &  \\ 
& -\frac 12(\tau ^{+})\int dx^4\varepsilon ^{\mu \nu \alpha \beta
}{}^{+}\Omega _{\mu \nu }^{\tau \lambda }{}^{+}G_{\alpha \beta }^{\sigma
\rho }{}\epsilon _{\tau \lambda \sigma \rho }+\frac 12(\tau ^{-})\int
dx^4\varepsilon ^{\mu \nu \alpha \beta }{}^{-}\Omega _{\mu \nu }^{\tau
\lambda }{}^{-}G_{\alpha \beta }^{\sigma \rho }{}\epsilon _{\tau \lambda
\sigma \rho } &  \\ 
&  &  \\ 
& +\frac 14(\tau ^{+})\int dx^4\varepsilon ^{\mu \nu \alpha \beta
}{}^{+}\Omega _{\mu \nu }^{\tau \lambda }{}^{+}\Omega _{\alpha \beta
}^{\sigma \rho }{}\epsilon _{\tau \lambda \sigma \rho }-\frac 14(\tau
^{-})\int dx^4\varepsilon ^{\mu \nu \alpha \beta }{}^{-}\Omega _{\mu \nu
}^{\tau \lambda }{}^{-}\Omega _{\alpha \beta }^{\sigma \rho }{}\epsilon
_{\tau \lambda \sigma \rho } &  \\ 
&  &  \\ 
& +\frac 12\int d^4x\varepsilon ^{\mu \nu \tau \lambda }{}^{+}W_{\mu \nu
}^{\alpha \beta }{}^{+}G_{\tau \lambda }^{\sigma \rho }{}\epsilon _{\alpha
\beta \sigma \rho }-\frac 12\int d^4x\varepsilon ^{\mu \nu \tau \lambda
}{}^{-}W_{\mu \nu }^{\alpha \beta }{}^{-}G_{\tau \lambda }^{\sigma \rho
}{}\epsilon _{\alpha \beta \sigma \rho }. & 
\end{array}
\label{32}
\end{equation}
Before integrating $h_{\mu \nu },$ let us consider the identity 
\begin{equation}
^{\pm }M_{\mu \nu }^{\tau \lambda }{}^{\pm }M_{\alpha \beta }^{\sigma \rho
}\epsilon _{\tau \lambda \sigma \rho }=\pm 4i^{\pm }M_{\mu \nu \alpha \beta }
\label{33}
\end{equation}
which can be obtained from the definition (22). Here, $^{\pm }M_{\mu \nu
\alpha \beta }=^{\pm }M_{\mu \nu }^{\tau \lambda }\eta _{\alpha \tau }\eta
_{\beta \lambda }.$ Using this identity and the definition (15) for $\Omega
_{\mu \nu }^{\tau \lambda }$ we obtain 
\begin{equation}
\begin{array}{c}
-\frac 12(\tau ^{+})\int d^4x\epsilon ^{\mu \nu \alpha \beta }{}^{+}\Omega
_{\mu \nu }^{\tau \lambda }{}^{+}G_{\alpha \beta }^{\sigma \rho }{}\epsilon
_{\tau \lambda \sigma \rho }+\frac 12(\tau ^{-})\int d^4x\epsilon ^{\mu \nu
\alpha \beta }{}^{-}\Omega _{\mu \nu }^{\tau \lambda }{}^{-}G_{\alpha \beta
}^{\sigma \rho }{}\epsilon _{\tau \lambda \sigma \rho } \\ 
\\ 
=\frac 12\int d^4x\epsilon ^{\mu \nu \alpha \beta }{}\Omega _{\mu \nu
}^{\tau \lambda }{}B_{\alpha \beta }^{\sigma \rho }{}\epsilon _{\tau \lambda
\sigma \rho }
\end{array}
\label{34}
\end{equation}
and 
\begin{equation}
\begin{array}{c}
\frac 14(\tau ^{+})\int d^4x\epsilon ^{\mu \nu \alpha \beta }{}^{+}\Omega
_{\mu \nu }^{\tau \lambda }{}^{+}\Omega _{\alpha \beta }^{\sigma \rho
}{}\epsilon _{\tau \lambda \sigma \rho }-\frac 14(\tau ^{-})\int
d^4x\epsilon ^{\mu \nu \alpha \beta }{}^{-}\Omega _{\mu \nu }^{\tau \lambda
}{}^{-}\Omega _{\alpha \beta }^{\sigma \rho }{}\epsilon _{\tau \lambda
\sigma \rho } \\ 
\\ 
=\frac b4\int d^4x\epsilon ^{\mu \nu \alpha \beta }{}\Omega _{\mu \nu
}^{\tau \lambda }{}\Omega _{\alpha \beta }^{\sigma \rho }{}\epsilon _{\tau
\lambda \sigma \rho },
\end{array}
\label{35}
\end{equation}
where 
\begin{equation}
B_{\alpha \beta }^{\sigma \rho }=-\{(\tau ^{+})^{+}G_{\alpha \beta }^{\sigma
\rho }-(\tau ^{-})^{-}G_{\alpha \beta }^{\sigma \rho }\}  \label{36}
\end{equation}
and 
\begin{equation}
b=\frac 12(\tau ^{+}-\tau ^{-}).  \label{37}
\end{equation}

Using (34)-(37) we learn that the action (32) can be written as 
\begin{equation}
\begin{array}{cc}
{\cal S}_E= & \frac 14(\tau ^{+})\int dx^4\varepsilon ^{\mu \nu \alpha \beta
}{}^{+}G_{\mu \nu }^{\tau \lambda }{}^{+}G_{\alpha \beta }^{\sigma \rho
}{}\epsilon _{\tau \lambda \sigma \rho }-\frac 14(\tau ^{-})\int
dx^4\varepsilon ^{\mu \nu \alpha \beta }{}^{-}G_{\mu \nu }^{\tau \lambda
}{}^{-}G_{\alpha \beta }^{\sigma \rho }{}\epsilon _{\tau \lambda \sigma \rho
} \\ 
&  \\ 
& +\frac 12\int d^4x\epsilon ^{\mu \nu \alpha \beta }{}\Omega _{\mu \nu
}^{\tau \lambda }{}B_{\alpha \beta }^{\sigma \rho }{}\epsilon _{\tau \lambda
\sigma \rho }+\frac b4\int d^4x\epsilon ^{\mu \nu \alpha \beta }{}\Omega
_{\mu \nu }^{\tau \lambda }{}\Omega _{\alpha \beta }^{\sigma \rho
}{}\epsilon _{\tau \lambda \sigma \rho } \\ 
&  \\ 
& +\frac 12\int d^4x\varepsilon ^{\mu \nu \tau \lambda }{}^{+}W_{\mu \nu
}^{\alpha \beta }{}^{+}G_{\tau \lambda }^{\sigma \rho }{}\epsilon _{\alpha
\beta \sigma \rho }-\frac 12\int d^4x\varepsilon ^{\mu \nu \tau \lambda
}{}^{-}W_{\mu \nu }^{\alpha \beta }{}^{-}G_{\tau \lambda }^{\sigma \rho
}{}\epsilon _{\alpha \beta \sigma \rho }.
\end{array}
\label{38}
\end{equation}
We note that the third and fourth term in (38) have exactly the same form as
the second and third term of (17). Therefore, using the definition (15) for $%
\Omega _{\mu \nu }^{\tau \lambda }$ and making similar calculations for
obtaining (19) we get 
\begin{equation}
\begin{array}{cc}
{\cal S}_E= & \frac 14(\tau ^{+})\int dx^4\varepsilon ^{\mu \nu \alpha \beta
}{}^{+}G_{\mu \nu }^{\tau \lambda }{}^{+}G_{\alpha \beta }^{\sigma \rho
}{}\epsilon _{\tau \lambda \sigma \rho }-\frac 14(\tau ^{-})\int
dx^4\varepsilon ^{\mu \nu \alpha \beta }{}^{-}G_{\mu \nu }^{\tau \lambda
}{}^{-}G_{\alpha \beta }^{\sigma \rho }{}\epsilon _{\tau \lambda \sigma \rho
} \\ 
&  \\ 
& +8\int d^4xh^{\mu \nu }(B_{\mu \nu }-\frac 12\eta _{\mu \nu }B)-8b\int
d^4x(h^2-h^{\mu \nu }h_{\mu \nu }) \\ 
&  \\ 
& +\frac 12\int d^4x\varepsilon ^{\mu \nu \tau \lambda }{}^{+}W_{\mu \nu
}^{\alpha \beta }{}^{+}G_{\tau \lambda }^{\sigma \rho }{}\epsilon _{\alpha
\beta \sigma \rho }-\frac 12\int d^4x\varepsilon ^{\mu \nu \tau \lambda
}{}^{-}W_{\mu \nu }^{\alpha \beta }{}^{-}G_{\tau \lambda }^{\sigma \rho
}{}\epsilon _{\alpha \beta \sigma \rho },
\end{array}
\label{39}
\end{equation}
where $B_{\mu \nu }\equiv \eta ^{\alpha \beta }B_{\mu \alpha \nu \beta },$
and $B\equiv \eta ^{\mu \nu }B_{\mu \nu }.$ Note that $b$ plays the role of
a cosmological constant.

We can now integrate out $h_{\mu \lambda }$ in the reduced partition
function 
\begin{equation}
Z=\int d^{+}G{}d^{-}G{}dh{}dV{}e^{-{\cal S}_E},  \label{40}
\end{equation}
with ${\cal S}_E$ given by (39). We first get the relation 
\begin{equation}
B_{\mu \nu }-\frac 12\eta _{\mu \nu }B=2b(\eta _{\mu \nu }h-h_{\mu \nu }).
\label{41}
\end{equation}
Using this expression the action ${\cal S}_E$ given in (39) becomes 
\begin{equation}
\begin{array}{cc}
{\cal S}_E= & \frac 14(\tau ^{+})\int d^4x\epsilon ^{\mu \nu \alpha \beta
}{}^{+}G_{\mu \nu }^{\tau \lambda }{}^{+}G_{\alpha \beta }^{\sigma \rho
}{}\epsilon _{\tau \lambda \sigma \rho }-\frac 14(\tau ^{-})\int
d^4x\epsilon ^{\mu \nu \alpha \beta }{}^{-}G_{\mu \nu }^{\tau \lambda
}{}^{-}G_{\alpha \beta }^{\sigma \rho }{}\epsilon _{\tau \lambda \sigma \rho
} \\ 
&  \\ 
& +\frac 2b\int d^4x(\frac 13B^2-B^{\mu \nu }B_{\mu \nu }) \\ 
&  \\ 
& +\frac 12\int d^4x\varepsilon ^{\mu \nu \tau \lambda }{}^{+}W_{\mu \nu
}^{\alpha \beta }{}^{+}G_{\tau \lambda }^{\sigma \rho }{}\epsilon _{\alpha
\beta \sigma \rho }-\frac 12\int d^4x\varepsilon ^{\mu \nu \tau \lambda
}{}^{-}W_{\mu \nu }^{\alpha \beta }{}^{-}G_{\tau \lambda }^{\sigma \rho
}{}\epsilon _{\alpha \beta \sigma \rho }.
\end{array}
\label{42}
\end{equation}
Before we perform the integrals over $^{+}G_{\mu \nu }^{\tau \lambda }$ and $%
^{-}G_{\mu \nu }^{\tau \lambda }$ we still need to rearrange the third
integral of (42). For this purpose let us introduce the definition 
\begin{equation}
B_{\mu \nu }=C_{\mu \nu }-\eta _{\mu \nu }C.  \label{43}
\end{equation}
Note first that $B=-3C.$ Therefore, we find $\frac 13B^2-B^{\mu \nu }B_{\mu
\nu }=C^2-C^{\mu \nu }C_{\mu \nu }.$ We now use the identity 
\begin{equation}
-\frac 1{32}\epsilon ^{\mu \nu \alpha \beta }{}S_{\mu \nu }^{\tau \lambda
}{}S_{\alpha \beta }^{\sigma \rho }{}\epsilon _{\tau \lambda \sigma \rho
}=(C^2-C^{\mu \nu }C_{\mu \nu }),  \label{44}
\end{equation}
where 
\begin{equation}
S_{\mu \nu }^{\alpha \beta }=\delta _\mu ^\alpha C_\nu ^\beta -\delta _\mu
^\beta C_\nu ^\alpha -\delta _\nu ^\alpha C_\mu ^\beta +\delta _\nu ^\beta
C_\mu ^\alpha  \label{45}
\end{equation}
and we observe that according to (36) we can write $S_{\mu \nu }^{\alpha
\beta }=-(\tau ^{+})_{+}S_{\mu \nu }^{\alpha \beta }+(\tau ^{-})_{-}S_{\mu
\nu }^{\alpha \beta },$ where

\begin{equation}
_{+}S_{\mu \nu }^{\alpha \beta }=\delta _\mu ^{\alpha +}\chi _\nu ^\beta
-\delta _\mu ^{\beta +}\chi _\nu ^\alpha -\delta _\nu ^{\alpha +}\chi _\mu
^\beta +\delta _\nu ^{\beta +}\chi _\mu ^\alpha  \label{46}
\end{equation}
and 
\begin{equation}
_{-}S_{\mu \nu }^{\alpha \beta }=\delta _\mu ^{\alpha -}\chi _\nu ^\beta
-\delta _\mu ^{\beta -}\chi _\nu ^\alpha -\delta _\nu ^{\alpha -}\chi _\mu
^\beta +\delta _\nu ^{\beta -}\chi _\mu ^\alpha .  \label{47}
\end{equation}
Here, $^{\pm }\chi _\nu ^\beta =G_{\nu \alpha }^{\pm \beta \alpha }-\frac 13%
\delta _\nu ^{\beta \pm }G_{\lambda \alpha }^{\lambda \alpha }$ and $C_{\mu
\nu }=-(\tau ^{+})^{+}\chi _{\mu \nu }+(\tau ^{-})^{-}\chi _{\mu \nu }.$
But, due to the symmetry properties of $G_{\mu \nu }^{\alpha \beta }$ given
in (24) we discover that $\chi _{\mu \nu }=^{+}\chi _{\mu \nu }=^{-}\chi
_{\mu \nu }$. So, $C_{\mu \nu }=-(\tau ^{+}-\tau ^{-})\chi _{\mu \nu }.$

Thus, we find that (42) can be written as 
\begin{equation}
\begin{array}{cc}
{\cal S}_E= & \frac 14(\tau ^{+})\int d^4x\epsilon ^{\mu \nu \alpha \beta
}{}^{+}G_{\mu \nu }^{\tau \lambda }{}^{+}G_{\alpha \beta }^{\sigma \rho
}{}\epsilon _{\tau \lambda \sigma \rho }-\frac 14(\tau ^{-})\int
d^4x\epsilon ^{\mu \nu \alpha \beta }{}^{-}G_{\mu \nu }^{\tau \lambda
}{}^{-}G_{\alpha \beta }^{\sigma \rho }{}\epsilon _{\tau \lambda \sigma \rho
} \\ 
&  \\ 
& -\frac 14(\tau ^{+})\int d^4x\epsilon ^{\mu \nu \alpha \beta }{}^{+}{\rm S}%
_{\mu \nu }^{\tau \lambda }{}^{+}{\rm S}_{\alpha \beta }^{\sigma \rho
}{}\epsilon _{\tau \lambda \sigma \rho }+\frac 14(\tau ^{-})\int
d^4x\epsilon ^{\mu \nu \alpha \beta }{}^{-}{\rm S}_{\mu \nu }^{\tau \lambda
}{}^{-}{\rm S}_{\alpha \beta }^{\sigma \rho }{}\epsilon _{\tau \lambda
\sigma \rho } \\ 
&  \\ 
& +\frac 12\int d^4x\varepsilon ^{\mu \nu \alpha \beta }{}^{+}W_{\mu \nu
}^{\tau \lambda }{}^{+}G_{\alpha \beta }^{\sigma \rho }{}\epsilon _{\tau
\lambda \sigma \rho }-\frac 12\int d^4x\varepsilon ^{\mu \nu \alpha \beta
}{}^{-}W_{\mu \nu }^{\tau \lambda }{}^{-}G_{\alpha \beta }^{\sigma \rho
}{}\epsilon _{\tau \lambda \sigma \rho }.
\end{array}
\label{48}
\end{equation}
Here, ${\rm S}_{\mu \nu }^{\alpha \beta }$ is defined as 
\begin{equation}
{\rm S}_{\mu \nu }^{\alpha \beta }=\delta _\mu ^\alpha \chi _\nu ^\beta
-\delta _\mu ^\beta \chi _\nu ^\alpha -\delta _\nu ^\alpha \chi _\mu ^\beta
+\delta _\nu ^\beta \chi _\mu ^\alpha .  \label{49}
\end{equation}

We are ready to perform the path integral of (48) with respect to $%
^{+}G_{\mu \nu }^{\tau \lambda }$ (similarly for $^{-}G_{\alpha \beta
}^{\sigma \rho }$). We get the constraint 
\begin{equation}
(\tau ^{+})^{+}G_{\mu \nu }^{\tau \lambda }+^{+}W_{\mu \nu }^{\tau \lambda }+%
\frac 14(\tau ^{+})\epsilon ^{\tau \lambda \sigma \rho }{}{\rm S}_{\sigma
\rho }^{\alpha \beta }\epsilon _{\alpha \beta \mu \nu }=0.  \label{50}
\end{equation}
It is not difficult to see that $(\tau ^{+})^{+}G_{\tau \nu }^{\tau \lambda
}+^{+}W_{\tau \nu }^{\tau \lambda }=0$ and $(\tau ^{+})^{+}G_{\tau \lambda
}^{\tau \lambda }+^{+}W_{\tau \lambda }^{\tau \lambda }=0.$ Using these two
expressions we find 
\begin{equation}
(\tau ^{+})^{+}G_{\mu \nu }^{\tau \lambda }+^{+}W_{\mu \nu }^{\tau \lambda }-%
\frac 14\epsilon ^{\tau \lambda \sigma \rho }{}{\rm L}_{\sigma \rho
}^{\alpha \beta }\epsilon _{\alpha \beta \mu \nu }=0,  \label{51}
\end{equation}
where 
\begin{equation}
{\rm L}_{\mu \nu }^{\alpha \beta }=\delta _\mu ^\alpha \zeta _\nu ^\beta
-\delta _\mu ^\beta \zeta _\nu ^\alpha -\delta _\nu ^\alpha \zeta _\mu
^\beta +\delta _\nu ^\beta \zeta _\mu ^\alpha ,  \label{52}
\end{equation}
with $^{\pm }\zeta _\nu ^\beta =W_{\nu \alpha }^{\pm \beta \alpha }-\frac 13%
\delta _\nu ^{\beta \pm }W_{\lambda \alpha }^{\lambda \alpha }.$ Now,
solving (51) for $^{+}G_{\mu \nu }^{\tau \lambda }$ and substituting the
result into (48) yields 
\begin{equation}
\begin{array}{cc}
S_E= & \frac 14(-\frac 1{\tau ^{+}})\int d^4x\epsilon ^{\mu \nu \alpha \beta
}{}^{+}W_{\mu \nu }^{\tau \lambda }{}^{+}W_{\alpha \beta }^{\sigma \rho
}{}\epsilon _{\tau \lambda \sigma \rho }-\frac 14(-\frac 1{\tau ^{-}})\int
d^4x\epsilon ^{\mu \nu \alpha \beta }{}^{-}W_{\mu \nu }^{\tau \lambda
}{}^{-}W_{\alpha \beta }^{\sigma \rho }{}\epsilon _{\tau \lambda \sigma \rho
} \\ 
&  \\ 
& -\frac 14(-\frac 1{\tau ^{+}})\int d^4x\epsilon ^{\mu \nu \alpha \beta
}{}^{+}{\rm L}_{\mu \nu }^{\tau \lambda }{}^{+}{\rm L}_{\alpha \beta
}^{\sigma \rho }{}\epsilon _{\tau \lambda \sigma \rho }+\frac 14(-\frac 1{%
\tau ^{-}})\int d^4x\epsilon ^{\mu \nu \alpha \beta }{}^{-}{\rm L}_{\mu \nu
}^{\tau \lambda }{}^{-}{\rm L}_{\alpha \beta }^{\sigma \rho }{}\epsilon
_{\tau \lambda \sigma \rho },
\end{array}
\label{53}
\end{equation}
where we apply similar procedure for $^{-}G_{\alpha \beta }^{\sigma \rho }$
. But, this action can be obtained from the following action: 
\begin{equation}
\begin{array}{cc}
S_D= & \frac 14(-\frac 1{\tau ^{+}})\int d^4x\epsilon ^{\mu \nu \alpha \beta
}{}^{+}Q_{\mu \nu }^{\tau \lambda }{}^{+}Q_{\alpha \beta }^{\sigma \rho
}{}\epsilon _{\tau \lambda \sigma \rho }-\frac 14(-\frac 1{\tau ^{-}})\int
d^4x\epsilon ^{\mu \nu \alpha \beta }{}^{-}Q_{\mu \nu }^{\tau \lambda
}{}^{-}Q_{\alpha \beta }^{\sigma \rho }{}\epsilon _{\tau \lambda \sigma \rho
},
\end{array}
\label{54}
\end{equation}
where 
\begin{equation}
^{\pm }Q_{\mu \nu }^{\tau \lambda }=^{\pm }W_{\mu \nu }^{\tau \lambda
}+^{\pm }{\rm K}_{\mu \nu }^{\tau \lambda },  \label{55}
\end{equation}
with 
\begin{equation}
{\rm K}_{\mu \nu }^{\alpha \beta }=\delta _\mu ^\alpha \gamma _\nu ^\beta
-\delta _\mu ^\beta \gamma _\nu ^\alpha -\delta _\nu ^\alpha \gamma _\mu
^\beta +\delta _\nu ^\beta \gamma _\mu ^\alpha .  \label{56}
\end{equation}
Here, $\gamma _{\mu \nu }$ is an auxiliary field. In fact, by integrating
out (54) with respect to $\gamma _{\mu \nu }$ we get (53). The action (54)
is, of course, the dual action. Note that the action (54) has the dual gauge
invariance $V\rightarrow V-d\alpha $ and is of the general type (20) but
with $\tau \rightarrow -\frac 1\tau ,$ where $\tau $ can be either $\tau
^{+} $ or $\tau ^{-}.$ Therefore, we have shown that the coupling transforms
as $\tau \rightarrow -\frac 1\tau $ when one changes from the action (20) to
the action (54)$.$ It is interesting to note that $\gamma _{\mu \nu }$ plays
the role of dual perturbation metric.

\smallskip\ 

\noindent {\bf 5.-FINAL COMMENTS}

\smallskip\ 

In this article, we have shown that it is possible to construct a S-dual
action for linearized gravity. Our two main ideas were to rewrite linearized
gravity as an Abelian gauge theory and to take recourse of
Macdowell-Mansouri formalism. The analysis of S-duality for a linearized
gravity then follows as in the case of Abelian gauge theories. An important
step was the realization that the transformation (28) provides with enough
symmetry to set $A=0$. After some long computations we finally prove that
the partition function for linearized gravity has the S-dual symmetry

\begin{equation}
Z_{A,h}(\tau )=Z_{V,\gamma }(-\frac 1\tau ),  \label{57}
\end{equation}
where $Z_{A,h}(\tau )$ is the partition function associated to the action
(20), while $Z_{V,\gamma }(-\frac 1\tau )$ is the partition function
associated to the action (54).

It seems that the present work can be extended without essential
complications to the case of linearized supergravity. As it is known, the
weak field limit of supergravity in four dimensions reduces to the sum of
the Fierz-Pauli and Rarita-Schwinger actions, which are the unique actions
for spin 2 and 3/2 without ghost (see [31] and references there in). This
means that we can write the partition function for linearized supergravity
as a product of two partition functions: one corresponding to linearized
gravity and the other corresponding to Rarita-Schwinger field. Taking
recourse of self-dual and antiself-dual supergravity and considering
separately self-dual-antiself-dual linearized gravity and
self-dual-antiself-dual Rarita-Shwinger, one can find the linearized
supergravity-S-dual symmetry . For linearized gravity one may proceed as in
previous sections, while for Rarita-Schwinger field one may use the method
developed in reference [11]. In this way, one should expect to get the
S-dual linearized supergravity symmetry $Z_{A,h,\psi }(\tau )=Z_{V,\gamma
,\varphi }(-\frac 1\tau ),$ where $\psi $ is the Rarita-Shwinger field and $%
\varphi $ its dual. At present, we are investigating in detail this
possibility and we expect to report our results in the near future.

It is known that the S-duality gauge invariance is a matter of great
interest in connection with M-theory [33-39] and supersymmetry [1,16,17] it
may be interesting to see whether the present work can be useful in those
directions.

Finally, it is worth mentioning the physical meaning, as well as the
possible relevance, of the strong coupling limit of linearized gravity. For
this purpose, we first need to clarify the physical meaning of the coupling
constants $\tau ^{+}$ and $\tau ^{-}.$ Write the action (16) as

\begin{equation}
S=\frac 1{16\Lambda }\int d^4x{}\epsilon ^{\mu \nu \alpha \beta }{}{\cal F}%
_{\mu \nu }^{\tau \lambda }{}{\cal F}_{\alpha \beta }^{\sigma \rho
}{}\epsilon _{\tau \lambda \sigma \rho }.  \label{58}
\end{equation}
As was mentioned at the end of the section 3, $\Lambda $ can be identified
with the cosmological constant. Now, using (22) it is not difficult to see
that the action (20) can be reduced to

\begin{equation}
{\cal S}=\frac 18(\tau ^{+}-\tau ^{-})\int d^4x{}\epsilon ^{\mu \nu \alpha
\beta }{}{\cal F}_{\mu \nu }^{\tau \lambda }{}{\cal F}_{\alpha \beta
}^{\sigma \rho }{}\epsilon _{\tau \lambda \sigma \rho }+\frac i8(\tau
^{+}+\tau ^{-})\int d^4x{}\epsilon ^{\mu \nu \alpha \beta }{}{\cal F}_{\mu
\nu }^{\tau \lambda }{}{\cal F}_{\alpha \beta }^{\sigma \rho }{}\delta
_{\tau \lambda \sigma \rho }.  \label{59}
\end{equation}
Thus, if we consider the expressions

\begin{eqnarray}
\tau ^{+} &=&\frac \Theta {2\pi }+\frac 1{4\Lambda },  \label{60} \\
\tau ^{-} &=&\frac \Theta {2\pi }-\frac 1{4\Lambda }  \nonumber
\end{eqnarray}
we find that the action (59) can be written as

\begin{equation}
{\cal S}=\frac 1{16\Lambda }\int d^4x{}\epsilon ^{\mu \nu \alpha \beta }{}%
{\cal F}_{\mu \nu }^{\tau \lambda }{}{\cal F}_{\alpha \beta }^{\sigma \rho
}{}\epsilon _{\tau \lambda \sigma \rho }+\frac{i\Theta }{8\pi }\int
d^4x{}\epsilon ^{\mu \nu \alpha \beta }{}{\cal F}_{\mu \nu }^{\tau \lambda
}{}{\cal F}_{\alpha \beta }^{\sigma \rho }{}\delta _{\tau \lambda \sigma
\rho }.  \label{61}
\end{equation}
If we now recall the case of gauge theory where $\tau ^{+}$ and $\tau ^{-}$
are defined as

\begin{eqnarray}
\tau ^{+} &=&\frac \theta {2\pi }+\frac{4\pi }{g^2},  \label{62} \\
\tau ^{-} &=&\frac \theta {2\pi }-\frac{4\pi }{g^2},  \nonumber
\end{eqnarray}
we see that the cosmological constant $\Lambda $ is playing the role the
gauge coupling constant $g^2$ and that $\Theta $ is playing the role of a $%
\theta $ constant. Just as in gauge theory the duality transformation $\tau
\rightarrow -\frac 1\tau ,$ where $\tau $ can be either $\tau ^{+}$ or $\tau
^{-},$ can be considered, in its simple form, as a duality of the gauge
coupling constant $g^2\rightarrow \frac 1{g^2}$ , in the case of linearized
gravity (according to (60)) the duality $\tau \rightarrow -\frac 1\tau $ can
be considered as a duality of the cosmological constant $\Lambda \rightarrow 
\frac 1\Lambda $. This result seems to suggest a new mechanism to make small
the cosmological constant, at least in linearized gravity (although similar
conclusions must be possible in more general cases as in references
[11]-[15]). In particular, the mechanism discussed in this work may be of
physical interest in Kaluza-Klein theories where in the process called
spontaneous compactifiction, normally, results a very large cosmological
constant.

\bigskip\ 

\noindent {\bf Acknowledgments}

\noindent This work was supported in part by CONACyT Grant 3898P-E9608.

\end{document}